\documentclass[showpacs, reprint]{revtex4-1}

\usepackage{amsthm, hyperref}

\begin{document}

\title{Scaling of quantum discord in spin models}

\author{Yichen Huang}

\email{yichenhuang@berkeley.edu}

\affiliation{Department of Physics, University of California, Berkeley, Berkeley, California 94720, USA}

\date{\today}

\begin{abstract}

We study the scaling of quantum discord (a measure of quantum correlation beyond entanglement) in spin models analytically and systematically. We find that at finite temperature the block scaling of quantum discord satisfies an area law for any two-local Hamiltonian. We show that generically and heuristically the two-site scaling of quantum discord is similar to that of correlation functions. In particular, at zero temperature it decays exponentially and polynomially in gapped and gapless (critical) systems, respectively; at finite temperature it decays exponentially in both gapped and gapless systems. We compute the two-site scaling of quantum discord in the $XXZ$ chain, the $XY$ chain (in a magnetic field), and the transverse field Ising chain at zero temperature.

\end{abstract}


\maketitle

Quite a few fundamental concepts in quantum mechanics do not have classical analogs: uncertainty relations \cite{Rob29, Bec75, BM75, Hua11, *Hua12}, quantum nonlocality \cite{EPR35, CHSH69, PV07, HHHH09}, etc. Quantum entanglement \cite{PV07, HHHH09}, defined based on the notion of local operations and classical communication, is the most prominent manifestation of quantum correlation. The set of separable (not entangled) states is convex and has nonzero measure (volume), and a lot of effort is devoted to entanglement detection \cite{HHHH09, Per96, DGCZ00, Sim00, GT09, Hua10, *Hua10E, *Hua13a}. However, nontrivial quantum correlation also exists in certain separable states. Quantum discord \cite{OZ01, HV01} is the most popular measure of quantum correlation beyond entanglement and an active research topic in the past few years \cite{MBC+12}. The set of classical (zero-discord) states is nowhere dense and has measure zero \cite{FAC+10}; computing quantum discord is difficult, since quantum discord is NP-complete.

Quantum discord has already been extensively studied in the context of condensed matter physics (spin chains) at both zero \cite{Dil08, Sar09, MGC+10} and finite \cite{MGC+10, WTRR10, WRR11} temperatures. These papers mainly focus on pairwise quantum discord between nearest-neighbor (or next-nearest-neighbor) spins and identify its singularities at quantum phase transitions. As emphasized in a recent review paper \cite{MBC+12}, the scaling of quantum discord is also an interesting problem. There are a few relevant numerical results \cite{MCSS12, CRGB13, CKG12}. The confidence that ``simple'' models (to some extent) describe complicated real materials comes from the notion of universality. The low-energy or long-range properties (the focus of condensed matter physics) are the same for different materials in the same universality class (though the high-energy or short-range properties are not), and models can be understood as representatives of universality classes. Therefore, the scaling of quantum discord may be more physically or practically relevant than pairwise quantum discord between nearest-neighbor spins.

We study the scaling of quantum discord in spin models analytically and systematically. We discuss two extensively studied notions of scaling in condensed matter physics: block scaling and two-site scaling. We find that at finite temperature the block scaling of quantum discord satisfies an area law for any two-local Hamiltonian. We show that generically and heuristically the two-site scaling of quantum discord is similar to that of correlation functions. In particular, at zero temperature it decays exponentially and polynomially in gapped and gapless (critical) systems, respectively; at finite temperature it decays exponentially in both gapped and gapless systems. We compute the two-site scaling of quantum discord in the $XXZ$ chain, the $XY$ chain (in a magnetic field), and the transverse field Ising chain at zero temperature.

\emph{Preliminaries.} In classical information theory, mutual information is the standard measure of correlation between two random variables. As the quantum analog, quantum mutual information
\begin{equation}
I(\rho_{AB})=S(\rho_A)+S(\rho_B)-S(\rho_{AB})
\end{equation}
quantifies the total (classical and quantum) correlation in a bipartite quantum state $\rho_{AB}$, where
\begin{equation}
S(\rho_A)=-\text{tr}\rho_A\ln\rho_A
\end{equation}
is the von Neumann entropy of the reduced density matrix $\rho_A=\mathrm{tr}_B\rho_{AB}$. When $\rho_{AB}$ is a pure state, $S(\rho_A)=S(\rho_B)$ is the so-called ``entanglement entropy'' \cite{PV07, HHHH09}. Entanglement entropy is the standard measure of entanglement for pure states, but not for mixed states.

Let $\{\Pi_i\}$ be a measurement on the subsystem $B$. Then, $p_i=\text{tr}(\Pi_i\rho_{AB})$ is the probability of the $i$th measurement outcome; $\rho_A^i=\text{tr}_B(\Pi_i\rho_{AB})/p_i$ and $\rho'_{AB}=\sum_ip_i\rho_A^i\otimes\Pi_i$ are the postmeasurement states. Classical correlation is defined as \cite{HV01}
\begin{equation}
J_B(\rho_{AB})=\max_{\{\Pi_i\}}J_{\{\Pi_i\}}(\rho_{AB})=\max_{\{\Pi_i\}}\left\{S(\rho_A)-\sum_ip_iS(\rho_A^i)\right\}.
\end{equation}
The maximization is taken either over all von Neumann measurements or over all generalized measurements described by positive-operator valued measures. For simplicity, we restrict ourselves to von Neumann measurements in this article. Quantum discord is the difference between total correlation and classical correlation \cite{OZ01},
\begin{eqnarray}
D_B(\rho_{AB})&=&I(\rho_{AB})-J_B(\rho_{AB})\nonumber\\
&=&\min_{\{\Pi_i\}}D_{\{\Pi_i\}}(\rho_{AB})=\min_{\{\Pi_i\}}\{I(\rho_{AB})-J_{\{\Pi_i\}}(\rho_{AB})\}\nonumber\\
&=&S(\rho_B)-S(\rho_{AB})+\min_{\{\Pi_i\}}\sum_ip_iS(\rho_A^i)\nonumber\\
&=&\min_{\{\Pi_i\}}S_B(\rho_{AB}')-S_B(\rho_{AB}),
\end{eqnarray}
where $S_B(\rho_{AB})=S(\rho_{AB})-S(\rho_B)$ is the quantum conditional entropy. Quantum discord reduces to entanglement entropy for pure states.

In physics, scaling typically refers to the asymptotic behavior of some quantity as some reference scale (e.g, length) diverges. Its mathematical definitions are different in different contexts. There are at least two extensively studied notions of scaling in condensed matter physics: block scaling and two-site scaling. Examples are, respectively, the asymptotic behavior of the entanglement entropy $S_L$ of a block of $L$ spins as $L\rightarrow+\infty$ and that of the correlation function $\langle X_iX_j\rangle$ as $|i-j|\rightarrow+\infty$ for some local operator $X$ at sites $i,j$.

\emph{Block scaling.} The block scaling of entanglement entropy and mutual information has been extensively studied in spin chains. At zero temperature the block scaling of entanglement entropy satisfies an area law \cite{ECP10} in one-dimensional (1D) gapped systems \cite{Has07}. The area law may be violated in 1D gapless (critical) systems. For instance, $S_L\sim (c\ln L)/3$, when the critical theory is a conformal field theory with central charge $c$ \cite{CC04, CC09}. At finite temperature the block scaling of mutual information satisfies an area law for any two-local Hamiltonian \cite{WVHC08}. Note that entanglement entropy is no longer a measure of entanglement for mixed states. 

The block scaling of quantum discord for a nondegenerate ground state is not new as quantum discord reduces to entanglement entropy for pure states. At finite temperature we observe a very general result that the block scaling of quantum discord satisfies an area law for any two-local Hamiltonian, regardless of the energy gap or the dimension/geometry of the underlying lattice. However, we believe that computing the block scaling of quantum discord numerically (or analytically) is difficult even for exactly solvable models.

\emph{Theorem 1.} (Area law for quantum discord) Let $G=(V,E)$ be the underlying lattice (graph) of a two-local Hamiltonian $H=\sum_{(i,j)\in E}h_{ij}$, where $V$ is the set of all sites in the spin system, and $h_{ij}|_{(i,j)\in E}$ acts only on the sites $i,j$. Let $A\subseteq V$ be a region and $B=V\backslash A$ be the rest. The thermal state $\rho_{AB}=e^{-\beta H}/\mathrm{tr}e^{-\beta H}$ at inverse temperature $\beta$ satisfies
\begin{equation}
D(\rho_{AB})\le2\beta|\partial A|\max_{(i,j)\in\partial A}\|h_{ij}\|_2,
\end{equation}
where $|\partial A|$ is the cardinality of the boundary $\partial A=\partial B=\{(i,j)\in E|i\in A,j\in B\}$, and $\|\cdot\|_p$ is the $p$-norm.

\begin{proof}
This is a straightforward consequence of the area law for mutual information \cite{WVHC08, ECP10} as mutual information is an upper bound on quantum discord. The thermal state $\rho_{AB}$ minimizes the free energy $F(\sigma)=\text{tr}(H\sigma)-S(\sigma)/\beta$. Therefore,
\begin{eqnarray}
F(\rho_{AB})&\le&F(\rho_A\otimes\rho_B)\Rightarrow\nonumber\\
D(\rho_{AB})&\le&I(\rho_{AB})\le\beta\sum_{(i,j)\in\partial A}\mathrm{tr}[h_{ij}(\rho_A\otimes\rho_B-\rho_{AB})]\nonumber\\
&\le&2\beta|\partial A|\max_{(i,j)\in\partial A}\|h_{ij}\|_2.
\end{eqnarray}
\end{proof}

\emph{Two-site scaling.} The two-site scaling of entanglement is generically trivial. In the absence of long-range order, the two-site reduced density matrix approaches the identity matrix $I$ as $|i-j|\rightarrow+\infty$. As the set of separable states contains a neighbourhood of $I$ \cite{GB02}, entanglement is exactly zero for not too small $|i-j|$ (for $|i-j|\gtrsim5$ in practice). This argument does not apply to the two-site scaling of quantum discord as the set of classical states is nowhere dense and has measure zero \cite{FAC+10}.

\emph{Theorem 2.} Let $\{X^k\}$ be a complete set of local operators and $\rho_\infty=\lim_{|i-j|\rightarrow+\infty}\rho_{ij}$. Suppose the convergence speed of all correlation functions is (tightly) bounded by a nonnegative monotonically decreasing function $f$:
\begin{equation}
|\langle X_i^{k_1}X_j^{k_2}\rangle-\langle X^{k_1}X^{k_2}\rangle_{\rho_\infty}|\le f(|i-j|)
\label{corr}
\end{equation}
for any $k_1,k_2$. Then,
\begin{equation}
|D(\rho_{ij})-D(\rho_\infty)|=O(-f(|i-j|)\ln f(|i-j|)).
\label{disc}
\end{equation}

\begin{proof}
As $\{X_i^{k_1}X_j^{k_2}\}$ is a complete set of two-site operators, Eq. (\ref{corr}) implies $\|\rho_{ij}-\rho_\infty\|_1=O(f(|i-j|))$. Then, Eq. (\ref{disc}) follows from the continuity of quantum discord \cite{XLWL11}
\begin{equation}
|D(\rho)-D(\sigma)|\le-4\|\rho-\sigma\|_1\ln\|\rho-\sigma\|_1+O(\|\rho-\sigma\|_1).
\end{equation}
\end{proof}

\emph{Remark.} Generically and heuristically, 
\begin{equation}
|D(\rho_{ij})-D(\rho_\infty)|\sim f^n(|i-j|),
\end{equation}
where $n$ (typically equal to 1,2) is a small positive integer. When the two-site scaling of quantum discord can be obtained from the scaling of correlation functions via Taylor expansion around $\rho_\infty$, $n$ is determined by the leading order in the expansion. Therefore, the two-site scaling of quantum discord is similar to that of correlation functions. In particular, at zero temperature it decays exponentially and polynomially in gapped and gapless (critical) systems, respectively; at finite temperature it decays exponentially in both gapped and gapless systems.

\emph{Methods.} In the remainder of this article, we explicitly compute the two-site scaling of quantum discord for the thermal ground state $\rho=\lim_{\beta\rightarrow+\infty}e^{-\beta H}/\mathrm{tr}e^{-\beta H}$ (which respects all the symmetries of the Hamiltonian $H$) in the $XXZ$ chain, the $XY$ chain (in a magnetic field), and the transverse field Ising chain. The $Z_2$ symmetry of these models implies that the two-site reduced density matrix $\rho_{ij}$ is a so-called ``X state'', which has nonzero elements only on the diagonal and the antidiagonal:
\begin{eqnarray}
\left[H,\bigotimes_{i}\sigma^z_i\right]=0&\Rightarrow&\left[\rho,\bigotimes_{i}\sigma^z_i\right]=0\nonumber\\
&\Rightarrow&[\rho_{ij},\sigma^z_i\otimes\sigma^z_j]=0\nonumber\\
&\Rightarrow&\rho_{ij}=\left(\begin{array}{cccc}a&0&0&\alpha\\0&b&\beta&0\\0&\bar\beta&c&0\\\bar\alpha&0&0&d\end{array}\right).
\end{eqnarray}
We emphasize that the analytical formula of quantum discord for general two-qubit X states is unknown (the formula claimed in \cite{ARA10,*ARA10E} is not exactly correct \cite{LMXW11, CZY+11, Hua13}). Fortunately, the following lemma suffices for our purpose.

\emph{Lemma 1 (\cite{CZY+11}).}
The optimal measurement is $\sigma^x$ if
\begin{equation}
|\sqrt{ad}-\sqrt{bc}|\le|\alpha|+|\beta|.
\label{cond}
\end{equation}

As in the computational basis the Hamiltonians are real, the matrix elements of $\rho_{ij}$ are all real, and are given by \cite{Sar09}
\begin{eqnarray}
&&a=(1+\langle\sigma^z_i\rangle+\langle\sigma^z_j\rangle+\langle\sigma^z_i\sigma^z_j\rangle)/4,\nonumber\\
&&b=(1+\langle\sigma^z_i\rangle-\langle\sigma^z_j\rangle-\langle\sigma^z_i\sigma^z_j\rangle)/4,\nonumber\\
&&c=(1-\langle\sigma^z_i\rangle+\langle\sigma^z_j\rangle-\langle\sigma^z_i\sigma^z_j\rangle)/4,\nonumber\\
&&d=(1-\langle\sigma^z_i\rangle-\langle\sigma^z_j\rangle+\langle\sigma^z_i\sigma^z_j\rangle)/4,\nonumber\\
&&\alpha=(\langle\sigma^x_i\sigma^x_j\rangle-\langle\sigma^y_i\sigma^y_j\rangle)/4,\nonumber\\
&&\beta=(\langle\sigma^x_i\sigma^x_j\rangle+\langle\sigma^y_i\sigma^y_j\rangle)/4.
\end{eqnarray}
The quantum conditional entropy is
\begin{equation}
S_j(\rho_{ij})=-H(1/2+\langle\sigma^z_i\rangle/2)-\sum_{k=1}^4\lambda_k\ln\lambda_k,
\label{qce}
\end{equation}
where $H(\lambda)=-\lambda\ln\lambda-(1-\lambda)\ln(1-\lambda)$ is the binary entropy function, and the eigenvalues of $\rho_{ij}$ are \cite{MCSS12}
\begin{eqnarray}
&&\lambda_{1,2}=(1-\langle\sigma^z_i\sigma^z_j\rangle\pm\langle\sigma^x_i\sigma^x_j\rangle\pm\langle\sigma^y_i\sigma^y_j\rangle)/4,\\
&&\lambda_{3,4}=\frac{1}{4}\left(1+\langle\sigma^z_i\sigma^z_j\rangle\pm\sqrt{4\langle\sigma^z_i\rangle^2+(\langle\sigma^x_i\sigma^x_j\rangle-\langle\sigma^y_i\sigma^y_j\rangle)^2}\right).
\end{eqnarray}
The quantum conditional entropy of the postmeasurement state $\rho'_{ij}$ after the (optimal) measurement $\sigma^x$ is
\begin{equation}
S_j(\rho'_{ij})=H\left(\frac{1}{2}+\frac{1}{2}\sqrt{\langle\sigma^z_i\rangle^2+\max\{\langle\sigma^x_i\sigma^x_j\rangle^2,\langle\sigma^y_i\sigma^y_j\rangle^2\}}\right).
\label{qcep}
\end{equation}

\emph{XXZ chain.} The Hamiltonian is
\begin{equation}
H=\sum_{i}\sigma^x_i\sigma^x_{i+1}+\sigma^y_i\sigma^y_{i+1}+\Delta\sigma^z_i\sigma^z_{i+1}.
\end{equation}
The model is exactly solvable by the Bethe ansatz. It is gapless (critical) for $-1<\Delta\le1$ and gapped for $|\Delta|>1$. In the critical regime $-1<\Delta<1$, the scaling of the correlation functions is \cite{LP75}
\begin{eqnarray}
&&\langle\sigma^z_i\sigma^z_j\rangle\sim A_z(-1)^{|i-j|}|i-j|^{-1/\eta}-\pi^{-2}\eta^{-1}|i-j|^{-2},\nonumber\\
&&\langle\sigma^x_i\sigma^x_j\rangle=\langle\sigma^y_i\sigma^y_j\rangle\sim A_x(-1)^{|i-j|}|i-j|^{-\eta},
\end{eqnarray}
where $0<\eta=\pi^{-1}\arccos(-\Delta)<1$, and the prefactors $A_z,A_x$ are given \cite{Luk97, HF98, Luk99} in the Appendix. The validity of (\ref{cond}) can be verified explicitly, and the optimal measurement is $\sigma^x$. We obtain the two-site scaling of quantum discord via the Taylor expansion of (\ref{qce}) and (\ref{qcep}):
\begin{equation}
D(\rho_{ij})\sim2^{-1}\langle\sigma^x_i\sigma^x_j\rangle^2\sim2^{-1}A_x^2|i-j|^{-2\eta}.
\end{equation}
At the critical point $\Delta=1$, the scaling of the correlation functions is \cite{Aff98, Luk98}
\begin{equation}
\langle\sigma^z_i\sigma^z_j\rangle=\langle\sigma^x_i\sigma^x_j\rangle=\langle\sigma^y_i\sigma^y_j\rangle\sim\frac{2^{1/2}(-1)^{|i-j|}\ln^{1/2}|i-j|}{\pi^{3/2}|i-j|}.
\end{equation}
As the antiferromagnetic Heisenberg model is SU(2) invariant, any measurement is optimal. We obtain the two-site scaling of quantum discord:
\begin{equation}
D(\rho_{ij})\sim\langle\sigma^x_i\sigma^x_j\rangle^2\sim2\pi^{-3}|i-j|^{-2}\ln|i-j|.
\end{equation}

\emph{XY chain.} The Hamiltonian is
\begin{equation}
H=-\sum_i\sigma^x_i\sigma^x_{i+1}+\alpha\sigma^y_i\sigma^y_{i+1}~(0\le\alpha\le1).
\end{equation}
The model is gapless (critical) for $\alpha=1$ \cite{LSM61} and gapped for $0\le\alpha<1$. In the gapped regime, the scaling of the correlation functions strongly depends on whether $i-j$ is even or odd \cite{Mcc68}. For even $i-j$,
\begin{eqnarray}
&&\langle\sigma^z_i\sigma^z_j\rangle=0,\nonumber\\
&&\langle\sigma^x_i\sigma^x_j\rangle\sim\sqrt{1-\alpha^2}+4\pi^{-1}(1-\alpha^2)^{-3/2}|i-j|^{-2}\alpha^{|i-j|+2},\nonumber\\
&&\langle\sigma^y_i\sigma^y_j\rangle\sim2\pi^{-1}(1-\alpha^2)^{-1/2}|i-j|^{-1}\alpha^{|i-j|};
\end{eqnarray}
and for odd $i-j$,
\begin{eqnarray}
&&\langle\sigma^z_i\sigma^z_j\rangle\sim-2\pi^{-1}|i-j|^{-2}\alpha^{|i-j|},\nonumber\\
&&\langle\sigma^x_i\sigma^x_j\rangle\sim\sqrt{1-\alpha^2}+\frac{2(1+\alpha^2)\alpha^{|i-j|+1}}{\pi(1-\alpha^2)^{3/2}|i-j|^2},\nonumber\\
&&\langle\sigma^y_i\sigma^y_j\rangle\sim2\pi^{-1}(1-\alpha^2)^{-1/2}|i-j|^{-1}\alpha^{|i-j|}.
\end{eqnarray}
The validity of (\ref{cond}) can be verified explicitly, and the optimal measurement is $\sigma^x$. We obtain the two-site scaling of quantum discord for both even and odd $i-j$:
\begin{eqnarray}
D(\rho_{ij})&\sim&2^{-1}\alpha^{-2}\langle\sigma^y_i\sigma^y_j\rangle^2\nonumber\\
&\sim&2\pi^{-2}(1-\alpha^2)^{-1}|i-j|^{-2}\alpha^{2|i-j|-2}.
\end{eqnarray} 

\emph{Transverse field Ising chain.} The Hamiltonian is
\begin{equation}
H=-\sum_i\sigma^x_i\sigma^x_{i+1}+h\sigma^z_i~(h\ge0).
\end{equation}
The model is gapless (critical) for $h=1$ and gapped otherwise. The magnetization is \cite{Pfe70}
\begin{equation}
\langle\sigma^z_i\rangle=\frac{1}{\pi}\int_0^\pi(h+\cos\theta)/\sqrt{(h+\cos\theta)^2+\sin^2\theta}\text{d}\theta,
\end{equation}
and in the ferromagnetic regime $h<1$ there is long-range order
\begin{equation}
\langle\sigma^x\sigma^x\rangle_\infty=\lim_{|i-j|\rightarrow+\infty}\langle\sigma^x_i\sigma^x_j\rangle=(1-h^2)^{1/4}.
\end{equation}
We only present the final results in order not to tire the reader with technical details. The two-site scaling of quantum discord is different in different regions:
\begin{eqnarray}
D(\rho_{ij})&\sim&A_1\langle\sigma^x_i\sigma^x_j\rangle^2\sim 2^{1/6}e^{1/2}A^{-6}A_1|i-j|^{-1/2},\\
D(\rho_{ij})&\sim&A_1\langle\sigma^x_i\sigma^x_j\rangle^2\sim\frac{A_1h^{-2|i-j|}}{\pi(1-h^{-2})^{1/2}|i-j|},\\
D(\rho_{ij})&\sim&D_{\sigma^x}(\rho_\infty)+A_2(\langle\sigma^x_i\sigma^x_j\rangle-\langle\sigma^x\sigma^x\rangle_\infty)\nonumber\\
&\sim&D_{\sigma^x}(\rho_\infty)+\frac{A_2h^{2|i-j|+2}}{2\pi(1-h^2)^{7/4}|i-j|^2}
\end{eqnarray}
for $h=1$, $h>1$, and $h<1$, respectively, where $A=1.2824271291\ldots$ is the Glaisher-Kinkelin constant, and the prefactors $A_1,A_2$ are given in the Appendix.

\emph{XY chain in a magnetic field.} The Hamiltonian is
\begin{equation}
H=-\sum_i(1+\gamma)\sigma^x_i\sigma^x_{i+1}/2+(1-\gamma)\sigma^y_i\sigma^y_{i+1}/2+h\sigma^z_i,
\end{equation}
where $0<\gamma\le1$ and $h\ge0$. The XY chain and the transverse field Ising chain are special cases of this model. This model is exactly solvable by the Jordan-Wigner transformation, and is essentially a model of free fermions. It is gapless (critical) for $h=1$ and gapped otherwise. The magnetization is \cite{Nie67}
\begin{equation}
\langle\sigma^z_i\rangle=\frac{1}{\pi}\int_0^\pi(h+\cos\theta)/\sqrt{(h+\cos\theta)^2+\gamma^2\sin^2\theta}\text{d}\theta,
\end{equation}
and in the regime $h<1$ there is long-range order \cite{BM71}
\begin{equation}
\langle\sigma^x\sigma^x\rangle_\infty=2\gamma^{1/2}(1+\gamma)^{-1}(1-h^2)^{1/4}.
\end{equation}
The two-site scaling of quantum discord is different in different regions:
\begin{eqnarray}
D(\rho_{ij})&\sim&A_1\langle\sigma^x_i\sigma^x_j\rangle^2\nonumber\\
&\sim&2^{13/6}e^{1/2}A^{-6}\gamma^{3/2}(1+\gamma)^{-2}A_1|i-j|^{-1/2},\label{XYm1}\\
D(\rho_{ij})&\sim&A_1\langle\sigma^x_i\sigma^x_j\rangle^2\nonumber\\
&\sim&\frac{2\gamma A_1\lambda^{2|i-j|}}{\pi(1+\gamma)^2|i-j|}\sqrt{1+\gamma^2+2\gamma\frac{1+\lambda^2}{1-\lambda^2}},\\
D(\rho_{ij})&\sim&D_{\sigma^x}(\rho_\infty)+A_2(\langle\sigma^x_i\sigma^x_j\rangle-\langle\sigma^x\sigma^x\rangle_\infty)\nonumber\\
&\sim&D_{\sigma^x}(\rho_\infty)+\frac{\gamma^{1/2}(1-h^2)^{1/4}A_2\lambda^{-2|i-j|}}{\pi(1+\gamma)(\lambda-\lambda^{-1})^2|i-j|^2},\\
D(\rho_{ij})&=&D_{\sigma^x}(\rho_\infty),\\
D(\rho_{ij})&\sim&D_{\sigma^x}(\rho_\infty)+A_3\langle\sigma^y_i\sigma^y_j\rangle\nonumber\\
&\sim&D_{\sigma^x}(\rho_\infty)+A_4|i-j|^{-1}[(1-\gamma)/(1+\gamma)]^{|i-j|}.\label{XYm2}
\end{eqnarray}
for $h=1$, $h>1$, $\sqrt{1-\gamma^2}<h<1$, $h=\sqrt{1-\gamma^2}$, and $0<h<\sqrt{1-\gamma^2}$, respectively, where
\begin{equation}
\lambda=(h-\sqrt{\gamma^2+h^2-1})/(1-\gamma),
\end{equation}
and the prefactors $A_3, A_4$ are given in the Appendix. The numerical results in \cite{MCSS12, CRGB13} are consistent with (\ref{XYm1})--(\ref{XYm2}).

\emph{Conclusion.} We have studied the scaling of quantum discord in spin models analytically and systematically. We have found that at finite temperature the block scaling of quantum discord satisfies an area law for any two-local Hamiltonian. We have shown that generically and heuristically the two-site scaling of quantum discord is similar to that of correlation functions. In particular, at zero temperature it decays exponentially and polynomially in gapped and gapless (critical) systems, respectively; at finite temperature it decays exponentially in both gapped and gapless systems. We have computed the two-site scaling of quantum discord in the $XXZ$ chain, the $XY$ chain (in a magnetic field), and the transverse field Ising chain at zero temperature.

\emph{Acknowledgments.} The author would like to thank Marcelo S. Sarandy for communications. This work was supported by ARO via the DARPA OLE program.

\appendix*

\section{}

The prefactors are given as follows:

\begin{widetext}

\begin{eqnarray}
A_z&=&2^{3-\frac{1}{\eta}}\pi^{-\frac{1}{2\eta}-2}\Gamma^{\frac{1}{\eta}}\left(\frac{\eta}{2-2\eta}\right)\Gamma^{-\frac{1}{\eta}}\left(\frac{1}{2-2\eta}\right)\exp\left[\int_0^{+\infty}\frac{\text dx}{x}\left(\frac{\sinh(2\eta x-x)}{\sinh(\eta x)\cosh(x-\eta x)}-\frac{2\eta-1}{\eta e^{2x}}\right)\right],\\
A_x&=&2^{-1-\eta}\pi^{-\frac{\eta}{2}}(1-\eta)^{-2}\Gamma^{\eta}\left(\frac{\eta}{2-2\eta}\right)\Gamma^{-\eta}\left(\frac{1}{2-2\eta}\right)\exp\left[\int_0^{+\infty}\frac{\text dx}{x}\left(\frac{\eta}{e^{2x}}-\frac{\sinh(\eta x)}{\sinh x\cosh(x-\eta x)}\right)\right],\\
A_1&=&4^{-1}(1-\langle\sigma^z_i\rangle^2)^{-1}+8^{-1}\langle\sigma^z_i\rangle^{-1}\ln[(1-\langle\sigma^z_i\rangle)/(1+\langle\sigma^z_i\rangle)],\\
A_2&=&\frac{1}{4}\ln\frac{1+\langle\sigma^x\sigma^x\rangle_\infty-\langle\sigma^z_i\rangle^2}{1-\langle\sigma^x\sigma^x\rangle_\infty-\langle\sigma^z_i\rangle^2}+\frac{\langle\sigma^x\sigma^x\rangle_\infty}{4\sqrt{\langle\sigma^x\sigma^x\rangle_\infty^2+4\langle\sigma^z_i\rangle^2}}\ln\frac{1+\langle\sigma^z_i\rangle^2+\sqrt{\langle\sigma^x\sigma^x\rangle_\infty^2+4\langle\sigma^z_i\rangle^2}}{1+\langle\sigma^z_i\rangle^2-\sqrt{\langle\sigma^x\sigma^x\rangle_\infty^2+4\langle\sigma^z_i\rangle^2}}\nonumber\\
&&+\frac{\langle\sigma^x\sigma^x\rangle_\infty}{2\sqrt{\langle\sigma^x\sigma^x\rangle_\infty^2+\langle\sigma^z_i\rangle^2}}\ln\frac{1-\sqrt{\langle\sigma^x\sigma^x\rangle_\infty^2+\langle\sigma^z_i\rangle^2}}{1+\sqrt{\langle\sigma^x\sigma^x\rangle_\infty^2+\langle\sigma^z_i\rangle^2}},\\
A_3&=&\frac{1}{4}\ln\frac{1+\langle\sigma^x\sigma^x\rangle_\infty-\langle\sigma^z_i\rangle^2}{1-\langle\sigma^x\sigma^x\rangle_\infty-\langle\sigma^z_i\rangle^2}+\frac{\langle\sigma^x\sigma^x\rangle_\infty}{4\sqrt{\langle\sigma^x\sigma^x\rangle_\infty^2+4\langle\sigma^z_i\rangle^2}}\ln\frac{1+\langle\sigma^z_i\rangle^2-\sqrt{\langle\sigma^x\sigma^x\rangle_\infty^2+4\langle\sigma^z_i\rangle^2}}{1+\langle\sigma^z_i\rangle^2+\sqrt{\langle\sigma^x\sigma^x\rangle_\infty^2+4\langle\sigma^z_i\rangle^2}},\\
A_4&=&4\pi^{-1}\gamma^{-1/2}(1-\gamma)^{-1}(1-h^2)^{1/4}(1-\gamma^2-h^2)|(1-\gamma)\lambda^2+(1+\gamma)\lambda^{-2}-2|^{-1}A_3.
\end{eqnarray}

\end{widetext}


\begin{thebibliography}{51}%
\makeatletter
\providecommand \@ifxundefined [1]{%
 \@ifx{#1\undefined}
}%
\providecommand \@ifnum [1]{%
 \ifnum #1\expandafter \@firstoftwo
 \else \expandafter \@secondoftwo
 \fi
}%
\providecommand \@ifx [1]{%
 \ifx #1\expandafter \@firstoftwo
 \else \expandafter \@secondoftwo
 \fi
}%
\providecommand \natexlab [1]{#1}%
\providecommand \enquote  [1]{``#1''}%
\providecommand \bibnamefont  [1]{#1}%
\providecommand \bibfnamefont [1]{#1}%
\providecommand \citenamefont [1]{#1}%
\providecommand \href@noop [0]{\@secondoftwo}%
\providecommand \href [0]{\begingroup \@sanitize@url \@href}%
\providecommand \@href[1]{\@@startlink{#1}\@@href}%
\providecommand \@@href[1]{\endgroup#1\@@endlink}%
\providecommand \@sanitize@url [0]{\catcode `\\12\catcode `\$12\catcode
  `\&12\catcode `\#12\catcode `\^12\catcode `\_12\catcode `\%12\relax}%
\providecommand \@@startlink[1]{}%
\providecommand \@@endlink[0]{}%
\providecommand \url  [0]{\begingroup\@sanitize@url \@url }%
\providecommand \@url [1]{\endgroup\@href {#1}{\urlprefix }}%
\providecommand \urlprefix  [0]{URL }%
\providecommand \Eprint [0]{\href }%
\providecommand \doibase [0]{http://dx.doi.org/}%
\providecommand \selectlanguage [0]{\@gobble}%
\providecommand \bibinfo  [0]{\@secondoftwo}%
\providecommand \bibfield  [0]{\@secondoftwo}%
\providecommand \translation [1]{[#1]}%
\providecommand \BibitemOpen [0]{}%
\providecommand \bibitemStop [0]{}%
\providecommand \bibitemNoStop [0]{.\EOS\space}%
\providecommand \EOS [0]{\spacefactor3000\relax}%
\providecommand \BibitemShut  [1]{\csname bibitem#1\endcsname}%
\let\auto@bib@innerbib\@empty
%</preamble>
\bibitem [{\citenamefont {Robertson}(1929)}]{Rob29}%
  \BibitemOpen
  \bibfield  {author} {\bibinfo {author} {\bibfnamefont {H.~P.}\ \bibnamefont
  {Robertson}},\ }\href {\doibase 10.1103/PhysRev.34.163} {\bibfield  {journal}
  {\bibinfo  {journal} {Phys. Rev.}\ }\textbf {\bibinfo {volume} {34}},\
  \bibinfo {pages} {163} (\bibinfo {year} {1929})}\BibitemShut {NoStop}%
\bibitem [{\citenamefont {Beckner}(1975)}]{Bec75}%
  \BibitemOpen
  \bibfield  {author} {\bibinfo {author} {\bibfnamefont {W.}~\bibnamefont
  {Beckner}},\ }\href {\doibase 10.2307/1970980} {\bibfield  {journal}
  {\bibinfo  {journal} {Ann. Math.}\ }\textbf {\bibinfo {volume} {102}},\
  \bibinfo {pages} {159} (\bibinfo {year} {1975})}\BibitemShut {NoStop}%
\bibitem [{\citenamefont {Bialynicki-Birula}\ and\ \citenamefont
  {Mycielski}(1975)}]{BM75}%
  \BibitemOpen
  \bibfield  {author} {\bibinfo {author} {\bibfnamefont {I.}~\bibnamefont
  {Bialynicki-Birula}}\ and\ \bibinfo {author} {\bibfnamefont {J.}~\bibnamefont
  {Mycielski}},\ }\href {\doibase 10.1007/BF01608825} {\bibfield  {journal}
  {\bibinfo  {journal} {Commun. Math. Phys.}\ }\textbf {\bibinfo {volume}
  {44}},\ \bibinfo {pages} {129} (\bibinfo {year} {1975})}\BibitemShut
  {NoStop}%
\bibitem [{\citenamefont {Huang}(2011)}]{Hua11}%
  \BibitemOpen
  \bibfield  {author} {\bibinfo {author} {\bibfnamefont {Y.}~\bibnamefont
  {Huang}},\ }\href {\doibase 10.1103/PhysRevA.83.052124} {\bibfield  {journal}
  {\bibinfo  {journal} {Phys. Rev. A}\ }\textbf {\bibinfo {volume} {83}},\
  \bibinfo {pages} {052124} (\bibinfo {year} {2011})}\BibitemShut {NoStop}%
\bibitem [{\citenamefont {Huang}(2012)}]{Hua12}%
  \BibitemOpen
  \bibfield  {author} {\bibinfo {author} {\bibfnamefont {Y.}~\bibnamefont
  {Huang}},\ }\href {\doibase 10.1103/PhysRevA.86.024101} {\bibfield  {journal}
  {\bibinfo  {journal} {Phys. Rev. A}\ }\textbf {\bibinfo {volume} {86}},\
  \bibinfo {pages} {024101} (\bibinfo {year} {2012})}\BibitemShut {NoStop}%
\bibitem [{\citenamefont {Einstein}\ \emph {et~al.}(1935)\citenamefont
  {Einstein}, \citenamefont {Podolsky},\ and\ \citenamefont {Rosen}}]{EPR35}%
  \BibitemOpen
  \bibfield  {author} {\bibinfo {author} {\bibfnamefont {A.}~\bibnamefont
  {Einstein}}, \bibinfo {author} {\bibfnamefont {B.}~\bibnamefont {Podolsky}},
  \ and\ \bibinfo {author} {\bibfnamefont {N.}~\bibnamefont {Rosen}},\ }\href
  {\doibase 10.1103/PhysRev.47.777} {\bibfield  {journal} {\bibinfo  {journal}
  {Phys. Rev.}\ }\textbf {\bibinfo {volume} {47}},\ \bibinfo {pages} {777}
  (\bibinfo {year} {1935})}\BibitemShut {NoStop}%
\bibitem [{\citenamefont {Clauser}\ \emph {et~al.}(1969)\citenamefont
  {Clauser}, \citenamefont {Horne}, \citenamefont {Shimony},\ and\
  \citenamefont {Holt}}]{CHSH69}%
  \BibitemOpen
  \bibfield  {author} {\bibinfo {author} {\bibfnamefont {J.~F.}\ \bibnamefont
  {Clauser}}, \bibinfo {author} {\bibfnamefont {M.~A.}\ \bibnamefont {Horne}},
  \bibinfo {author} {\bibfnamefont {A.}~\bibnamefont {Shimony}}, \ and\
  \bibinfo {author} {\bibfnamefont {R.~A.}\ \bibnamefont {Holt}},\ }\href
  {\doibase 10.1103/PhysRevLett.23.880} {\bibfield  {journal} {\bibinfo
  {journal} {Phys. Rev. Lett.}\ }\textbf {\bibinfo {volume} {23}},\ \bibinfo
  {pages} {880} (\bibinfo {year} {1969})}\BibitemShut {NoStop}%
\bibitem [{\citenamefont {Plenio}\ and\ \citenamefont {Virmani}(2007)}]{PV07}%
  \BibitemOpen
  \bibfield  {author} {\bibinfo {author} {\bibfnamefont {M.~B.}\ \bibnamefont
  {Plenio}}\ and\ \bibinfo {author} {\bibfnamefont {S.}~\bibnamefont
  {Virmani}},\ }\href@noop {} {\bibfield  {journal} {\bibinfo  {journal}
  {Quantum Inf. Comput.}\ }\textbf {\bibinfo {volume} {7}},\ \bibinfo {pages}
  {1} (\bibinfo {year} {2007})}\BibitemShut {NoStop}%
\bibitem [{\citenamefont {Horodecki}\ \emph {et~al.}(2009)\citenamefont
  {Horodecki}, \citenamefont {Horodecki}, \citenamefont {Horodecki},\ and\
  \citenamefont {Horodecki}}]{HHHH09}%
  \BibitemOpen
  \bibfield  {author} {\bibinfo {author} {\bibfnamefont {R.}~\bibnamefont
  {Horodecki}}, \bibinfo {author} {\bibfnamefont {P.}~\bibnamefont
  {Horodecki}}, \bibinfo {author} {\bibfnamefont {M.}~\bibnamefont
  {Horodecki}}, \ and\ \bibinfo {author} {\bibfnamefont {K.}~\bibnamefont
  {Horodecki}},\ }\href {\doibase 10.1103/RevModPhys.81.865} {\bibfield
  {journal} {\bibinfo  {journal} {Rev. Mod. Phys.}\ }\textbf {\bibinfo {volume}
  {81}},\ \bibinfo {pages} {865} (\bibinfo {year} {2009})}\BibitemShut
  {NoStop}%
\bibitem [{\citenamefont {Peres}(1996)}]{Per96}%
  \BibitemOpen
  \bibfield  {author} {\bibinfo {author} {\bibfnamefont {A.}~\bibnamefont
  {Peres}},\ }\href {\doibase 10.1103/PhysRevLett.77.1413} {\bibfield
  {journal} {\bibinfo  {journal} {Phys. Rev. Lett.}\ }\textbf {\bibinfo
  {volume} {77}},\ \bibinfo {pages} {1413} (\bibinfo {year}
  {1996})}\BibitemShut {NoStop}%
\bibitem [{\citenamefont {Duan}\ \emph {et~al.}(2000)\citenamefont {Duan},
  \citenamefont {Giedke}, \citenamefont {Cirac},\ and\ \citenamefont
  {Zoller}}]{DGCZ00}%
  \BibitemOpen
  \bibfield  {author} {\bibinfo {author} {\bibfnamefont {L.-M.}\ \bibnamefont
  {Duan}}, \bibinfo {author} {\bibfnamefont {G.}~\bibnamefont {Giedke}},
  \bibinfo {author} {\bibfnamefont {J.~I.}\ \bibnamefont {Cirac}}, \ and\
  \bibinfo {author} {\bibfnamefont {P.}~\bibnamefont {Zoller}},\ }\href
  {\doibase 10.1103/PhysRevLett.84.2722} {\bibfield  {journal} {\bibinfo
  {journal} {Phys. Rev. Lett.}\ }\textbf {\bibinfo {volume} {84}},\ \bibinfo
  {pages} {2722} (\bibinfo {year} {2000})}\BibitemShut {NoStop}%
\bibitem [{\citenamefont {Simon}(2000)}]{Sim00}%
  \BibitemOpen
  \bibfield  {author} {\bibinfo {author} {\bibfnamefont {R.}~\bibnamefont
  {Simon}},\ }\href {\doibase 10.1103/PhysRevLett.84.2726} {\bibfield
  {journal} {\bibinfo  {journal} {Phys. Rev. Lett.}\ }\textbf {\bibinfo
  {volume} {84}},\ \bibinfo {pages} {2726} (\bibinfo {year}
  {2000})}\BibitemShut {NoStop}%
\bibitem [{\citenamefont {Guhne}\ and\ \citenamefont {Toth}(2009)}]{GT09}%
  \BibitemOpen
  \bibfield  {author} {\bibinfo {author} {\bibfnamefont {O.}~\bibnamefont
  {Guhne}}\ and\ \bibinfo {author} {\bibfnamefont {G.}~\bibnamefont {Toth}},\
  }\href {\doibase 10.1016/j.physrep.2009.02.004} {\bibfield  {journal}
  {\bibinfo  {journal} {Phys. Rep.}\ }\textbf {\bibinfo {volume} {474}},\
  \bibinfo {pages} {1 } (\bibinfo {year} {2009})}\BibitemShut {NoStop}%
\bibitem [{\citenamefont {Huang}(2010{\natexlab{a}})}]{Hua10}%
  \BibitemOpen
  \bibfield  {author} {\bibinfo {author} {\bibfnamefont {Y.}~\bibnamefont
  {Huang}},\ }\href {\doibase 10.1103/PhysRevA.82.012335} {\bibfield  {journal}
  {\bibinfo  {journal} {Phys. Rev. A}\ }\textbf {\bibinfo {volume} {82}},\
  \bibinfo {pages} {012335} (\bibinfo {year} {2010}{\natexlab{a}})}\BibitemShut
  {NoStop}%
\bibitem [{\citenamefont {Huang}(2010{\natexlab{b}})}]{Hua10E}%
  \BibitemOpen
  \bibfield  {author} {\bibinfo {author} {\bibfnamefont {Y.}~\bibnamefont
  {Huang}},\ }\href {\doibase 10.1103/PhysRevA.82.069903} {\bibfield  {journal}
  {\bibinfo  {journal} {Phys. Rev. A}\ }\textbf {\bibinfo {volume} {82}},\
  \bibinfo {pages} {069903} (\bibinfo {year} {2010}{\natexlab{b}})}\BibitemShut
  {NoStop}%
\bibitem [{\citenamefont {Huang}(2013{\natexlab{a}})}]{Hua13a}%
  \BibitemOpen
  \bibfield  {author} {\bibinfo {author} {\bibfnamefont {Y.}~\bibnamefont
  {Huang}},\ }\href {\doibase 10.1109/TIT.2013.2257936} {\bibfield  {journal}
  {\bibinfo  {journal} {IEEE Trans. Inf. Theory}\ }\textbf {\bibinfo {volume}
  {59}},\ \bibinfo {pages} {6774} (\bibinfo {year}
  {2013}{\natexlab{a}})}\BibitemShut {NoStop}%
\bibitem [{\citenamefont {Ollivier}\ and\ \citenamefont {Zurek}(2001)}]{OZ01}%
  \BibitemOpen
  \bibfield  {author} {\bibinfo {author} {\bibfnamefont {H.}~\bibnamefont
  {Ollivier}}\ and\ \bibinfo {author} {\bibfnamefont {W.~H.}\ \bibnamefont
  {Zurek}},\ }\href {\doibase 10.1103/PhysRevLett.88.017901} {\bibfield
  {journal} {\bibinfo  {journal} {Phys. Rev. Lett.}\ }\textbf {\bibinfo
  {volume} {88}},\ \bibinfo {pages} {017901} (\bibinfo {year}
  {2001})}\BibitemShut {NoStop}%
\bibitem [{\citenamefont {Henderson}\ and\ \citenamefont
  {Vedral}(2001)}]{HV01}%
  \BibitemOpen
  \bibfield  {author} {\bibinfo {author} {\bibfnamefont {L.}~\bibnamefont
  {Henderson}}\ and\ \bibinfo {author} {\bibfnamefont {V.}~\bibnamefont
  {Vedral}},\ }\href {\doibase 10.1088/0305-4470/34/35/315} {\bibfield
  {journal} {\bibinfo  {journal} {J. Phys. A: Math. Gen.}\ }\textbf {\bibinfo
  {volume} {34}},\ \bibinfo {pages} {6899} (\bibinfo {year}
  {2001})}\BibitemShut {NoStop}%
\bibitem [{\citenamefont {Modi}\ \emph {et~al.}(2012)\citenamefont {Modi},
  \citenamefont {Brodutch}, \citenamefont {Cable}, \citenamefont {Paterek},\
  and\ \citenamefont {Vedral}}]{MBC+12}%
  \BibitemOpen
  \bibfield  {author} {\bibinfo {author} {\bibfnamefont {K.}~\bibnamefont
  {Modi}}, \bibinfo {author} {\bibfnamefont {A.}~\bibnamefont {Brodutch}},
  \bibinfo {author} {\bibfnamefont {H.}~\bibnamefont {Cable}}, \bibinfo
  {author} {\bibfnamefont {T.}~\bibnamefont {Paterek}}, \ and\ \bibinfo
  {author} {\bibfnamefont {V.}~\bibnamefont {Vedral}},\ }\href {\doibase 10.1103/RevModPhys.84.1655} {\bibfield  {journal} {\bibinfo  {journal} {Rev.
  Mod. Phys.}\ }\textbf {\bibinfo {volume} {84}},\ \bibinfo {pages} {1655}
  (\bibinfo {year} {2012})}\BibitemShut {NoStop}%
\bibitem [{\citenamefont {Ferraro}\ \emph {et~al.}(2010)\citenamefont
  {Ferraro}, \citenamefont {Aolita}, \citenamefont {Cavalcanti}, \citenamefont
  {Cucchietti},\ and\ \citenamefont {Acin}}]{FAC+10}%
  \BibitemOpen
  \bibfield  {author} {\bibinfo {author} {\bibfnamefont {A.}~\bibnamefont
  {Ferraro}}, \bibinfo {author} {\bibfnamefont {L.}~\bibnamefont {Aolita}},
  \bibinfo {author} {\bibfnamefont {D.}~\bibnamefont {Cavalcanti}}, \bibinfo
  {author} {\bibfnamefont {F.~M.}\ \bibnamefont {Cucchietti}}, \ and\ \bibinfo
  {author} {\bibfnamefont {A.}~\bibnamefont {Acin}},\ }\href {\doibase 10.1103/PhysRevA.81.052318} {\bibfield  {journal} {\bibinfo  {journal} {Phys.
  Rev. A}\ }\textbf {\bibinfo {volume} {81}},\ \bibinfo {pages} {052318}
  (\bibinfo {year} {2010})}\BibitemShut {NoStop}%
\bibitem [{\citenamefont {Dillenschneider}(2008)}]{Dil08}%
  \BibitemOpen
  \bibfield  {author} {\bibinfo {author} {\bibfnamefont {R.}~\bibnamefont
  {Dillenschneider}},\ }\href {\doibase 10.1103/PhysRevB.78.224413} {\bibfield
  {journal} {\bibinfo  {journal} {Phys. Rev. B}\ }\textbf {\bibinfo {volume}
  {78}},\ \bibinfo {pages} {224413} (\bibinfo {year} {2008})}\BibitemShut
  {NoStop}%
\bibitem [{\citenamefont {Sarandy}(2009)}]{Sar09}%
  \BibitemOpen
  \bibfield  {author} {\bibinfo {author} {\bibfnamefont {M.~S.}\ \bibnamefont
  {Sarandy}},\ }\href {\doibase 10.1103/PhysRevA.80.022108} {\bibfield
  {journal} {\bibinfo  {journal} {Phys. Rev. A}\ }\textbf {\bibinfo {volume}
  {80}},\ \bibinfo {pages} {022108} (\bibinfo {year} {2009})}\BibitemShut
  {NoStop}%
\bibitem [{\citenamefont {Maziero}\ \emph {et~al.}(2010)\citenamefont
  {Maziero}, \citenamefont {Guzman}, \citenamefont {Celeri}, \citenamefont
  {Sarandy},\ and\ \citenamefont {Serra}}]{MGC+10}%
  \BibitemOpen
  \bibfield  {author} {\bibinfo {author} {\bibfnamefont {J.}~\bibnamefont
  {Maziero}}, \bibinfo {author} {\bibfnamefont {H.~C.}\ \bibnamefont {Guzman}},
  \bibinfo {author} {\bibfnamefont {L.~C.}\ \bibnamefont {Celeri}}, \bibinfo
  {author} {\bibfnamefont {M.~S.}\ \bibnamefont {Sarandy}}, \ and\ \bibinfo
  {author} {\bibfnamefont {R.~M.}\ \bibnamefont {Serra}},\ }\href {\doibase 10.1103/PhysRevA.82.012106} {\bibfield  {journal} {\bibinfo  {journal} {Phys.
  Rev. A}\ }\textbf {\bibinfo {volume} {82}},\ \bibinfo {pages} {012106}
  (\bibinfo {year} {2010})}\BibitemShut {NoStop}%
\bibitem [{\citenamefont {Werlang}\ \emph {et~al.}(2010)\citenamefont
  {Werlang}, \citenamefont {Trippe}, \citenamefont {Ribeiro},\ and\
  \citenamefont {Rigolin}}]{WTRR10}%
  \BibitemOpen
  \bibfield  {author} {\bibinfo {author} {\bibfnamefont {T.}~\bibnamefont
  {Werlang}}, \bibinfo {author} {\bibfnamefont {C.}~\bibnamefont {Trippe}},
  \bibinfo {author} {\bibfnamefont {G.~A.~P.}\ \bibnamefont {Ribeiro}}, \ and\
  \bibinfo {author} {\bibfnamefont {G.}~\bibnamefont {Rigolin}},\ }\href
  {\doibase 10.1103/PhysRevLett.105.095702} {\bibfield  {journal} {\bibinfo
  {journal} {Phys. Rev. Lett.}\ }\textbf {\bibinfo {volume} {105}},\ \bibinfo
  {pages} {095702} (\bibinfo {year} {2010})}\BibitemShut {NoStop}%
\bibitem [{\citenamefont {Werlang}\ \emph {et~al.}(2011)\citenamefont
  {Werlang}, \citenamefont {Ribeiro},\ and\ \citenamefont {Rigolin}}]{WRR11}%
  \BibitemOpen
  \bibfield  {author} {\bibinfo {author} {\bibfnamefont {T.}~\bibnamefont
  {Werlang}}, \bibinfo {author} {\bibfnamefont {G.~A.~P.}\ \bibnamefont
  {Ribeiro}}, \ and\ \bibinfo {author} {\bibfnamefont {G.}~\bibnamefont
  {Rigolin}},\ }\href {\doibase 10.1103/PhysRevA.83.062334} {\bibfield
  {journal} {\bibinfo  {journal} {Phys. Rev. A}\ }\textbf {\bibinfo {volume}
  {83}},\ \bibinfo {pages} {062334} (\bibinfo {year} {2011})}\BibitemShut
  {NoStop}%
\bibitem [{\citenamefont {Maziero}\ \emph {et~al.}(2012)\citenamefont
  {Maziero}, \citenamefont {Celeri}, \citenamefont {Serra},\ and\ \citenamefont
  {Sarandy}}]{MCSS12}%
  \BibitemOpen
  \bibfield  {author} {\bibinfo {author} {\bibfnamefont {J.}~\bibnamefont
  {Maziero}}, \bibinfo {author} {\bibfnamefont {L.}~\bibnamefont {Celeri}},
  \bibinfo {author} {\bibfnamefont {R.}~\bibnamefont {Serra}}, \ and\ \bibinfo
  {author} {\bibfnamefont {M.}~\bibnamefont {Sarandy}},\ }\href {\doibase 10.1016/j.physleta.2012.03.029} {\bibfield  {journal} {\bibinfo  {journal}
  {Phys. Lett. A}\ }\textbf {\bibinfo {volume} {376}},\ \bibinfo {pages} {1540
  } (\bibinfo {year} {2012})}\BibitemShut {NoStop}%
\bibitem [{\citenamefont {Campbell}\ \emph {et~al.}(2013)\citenamefont
  {Campbell}, \citenamefont {Richens}, \citenamefont {Gullo},\ and\
  \citenamefont {Busch}}]{CRGB13}%
  \BibitemOpen
  \bibfield  {author} {\bibinfo {author} {\bibfnamefont {S.}~\bibnamefont
  {Campbell}}, \bibinfo {author} {\bibfnamefont {J.}~\bibnamefont {Richens}},
  \bibinfo {author} {\bibfnamefont {N.~L.}\ \bibnamefont {Gullo}}, \ and\
  \bibinfo {author} {\bibfnamefont {T.}~\bibnamefont {Busch}},\ }\href
  {\doibase 10.1103/PhysRevA.88.062305} {\bibfield  {journal} {\bibinfo
  {journal} {Phys. Rev. A}\ }\textbf {\bibinfo {volume} {88}},\ \bibinfo
  {pages} {062305} (\bibinfo {year} {2013})}\BibitemShut {NoStop}%
\bibitem [{\citenamefont {Cakmak}\ \emph {et~al.}(2012)\citenamefont {Cakmak},
  \citenamefont {Karpat},\ and\ \citenamefont {Gedik}}]{CKG12}%
  \BibitemOpen
  \bibfield  {author} {\bibinfo {author} {\bibfnamefont {B.}~\bibnamefont
  {Cakmak}}, \bibinfo {author} {\bibfnamefont {G.}~\bibnamefont {Karpat}}, \
  and\ \bibinfo {author} {\bibfnamefont {Z.}~\bibnamefont {Gedik}},\ }\href
  {\doibase 10.1016/j.physleta.2012.09.007} {\bibfield  {journal} {\bibinfo
  {journal} {Phys. Lett. A}\ }\textbf {\bibinfo {volume} {376}},\ \bibinfo
  {pages} {2982} (\bibinfo {year} {2012})}\BibitemShut {NoStop}%
\bibitem [{\citenamefont {Eisert}\ \emph {et~al.}(2010)\citenamefont {Eisert},
  \citenamefont {Cramer},\ and\ \citenamefont {Plenio}}]{ECP10}%
  \BibitemOpen
  \bibfield  {author} {\bibinfo {author} {\bibfnamefont {J.}~\bibnamefont
  {Eisert}}, \bibinfo {author} {\bibfnamefont {M.}~\bibnamefont {Cramer}}, \
  and\ \bibinfo {author} {\bibfnamefont {M.~B.}\ \bibnamefont {Plenio}},\
  }\href {\doibase 10.1103/RevModPhys.82.277} {\bibfield  {journal} {\bibinfo
  {journal} {Rev. Mod. Phys.}\ }\textbf {\bibinfo {volume} {82}},\ \bibinfo
  {pages} {277} (\bibinfo {year} {2010})}\BibitemShut {NoStop}%
\bibitem [{\citenamefont {Hastings}(2007)}]{Has07}%
  \BibitemOpen
  \bibfield  {author} {\bibinfo {author} {\bibfnamefont {M.~B.}\ \bibnamefont
  {Hastings}},\ }\href {\doibase 10.1088/1742-5468/2007/08/P08024} {\bibfield
  {journal} {\bibinfo  {journal} {J. Stat. Mech.}\ }\textbf {\bibinfo {volume}
  {2007}},\ \bibinfo {pages} {P08024} (\bibinfo {year} {2007})}\BibitemShut
  {NoStop}%
\bibitem [{\citenamefont {Calabrese}\ and\ \citenamefont {Cardy}(2004)}]{CC04}%
  \BibitemOpen
  \bibfield  {author} {\bibinfo {author} {\bibfnamefont {P.}~\bibnamefont
  {Calabrese}}\ and\ \bibinfo {author} {\bibfnamefont {J.}~\bibnamefont
  {Cardy}},\ }\href {\doibase 10.1088/1742-5468/2004/06/P06002} {\bibfield
  {journal} {\bibinfo  {journal} {J. Stat. Mech.}\ }\textbf {\bibinfo {volume}
  {2004}},\ \bibinfo {pages} {P06002} (\bibinfo {year} {2004})}\BibitemShut
  {NoStop}%
\bibitem [{\citenamefont {Calabrese}\ and\ \citenamefont {Cardy}(2009)}]{CC09}%
  \BibitemOpen
  \bibfield  {author} {\bibinfo {author} {\bibfnamefont {P.}~\bibnamefont
  {Calabrese}}\ and\ \bibinfo {author} {\bibfnamefont {J.}~\bibnamefont
  {Cardy}},\ }\href {\doibase 10.1088/1751-8113/42/50/504005} {\bibfield
  {journal} {\bibinfo  {journal} {J. Phys. A: Math. Theor.}\ }\textbf {\bibinfo
  {volume} {42}},\ \bibinfo {pages} {504005} (\bibinfo {year}
  {2009})}\BibitemShut {NoStop}%
\bibitem [{\citenamefont {Wolf}\ \emph {et~al.}(2008)\citenamefont {Wolf},
  \citenamefont {Verstraete}, \citenamefont {Hastings},\ and\ \citenamefont
  {Cirac}}]{WVHC08}%
  \BibitemOpen
  \bibfield  {author} {\bibinfo {author} {\bibfnamefont {M.~M.}\ \bibnamefont
  {Wolf}}, \bibinfo {author} {\bibfnamefont {F.}~\bibnamefont {Verstraete}},
  \bibinfo {author} {\bibfnamefont {M.~B.}\ \bibnamefont {Hastings}}, \ and\
  \bibinfo {author} {\bibfnamefont {J.~I.}\ \bibnamefont {Cirac}},\ }\href
  {\doibase 10.1103/PhysRevLett.100.070502} {\bibfield  {journal} {\bibinfo
  {journal} {Phys. Rev. Lett.}\ }\textbf {\bibinfo {volume} {100}},\ \bibinfo
  {pages} {070502} (\bibinfo {year} {2008})}\BibitemShut {NoStop}%
\bibitem [{\citenamefont {Gurvits}\ and\ \citenamefont {Barnum}(2002)}]{GB02}%
  \BibitemOpen
  \bibfield  {author} {\bibinfo {author} {\bibfnamefont {L.}~\bibnamefont
  {Gurvits}}\ and\ \bibinfo {author} {\bibfnamefont {H.}~\bibnamefont
  {Barnum}},\ }\href {\doibase 10.1103/PhysRevA.66.062311} {\bibfield
  {journal} {\bibinfo  {journal} {Phys. Rev. A}\ }\textbf {\bibinfo {volume}
  {66}},\ \bibinfo {pages} {062311} (\bibinfo {year} {2002})}\BibitemShut
  {NoStop}%
\bibitem [{\citenamefont {Xi}\ \emph {et~al.}(2011)\citenamefont {Xi},
  \citenamefont {Lu}, \citenamefont {Wang},\ and\ \citenamefont {Li}}]{XLWL11}%
  \BibitemOpen
  \bibfield  {author} {\bibinfo {author} {\bibfnamefont {Z.}~\bibnamefont
  {Xi}}, \bibinfo {author} {\bibfnamefont {X.-M.}\ \bibnamefont {Lu}}, \bibinfo
  {author} {\bibfnamefont {X.}~\bibnamefont {Wang}}, \ and\ \bibinfo {author}
  {\bibfnamefont {Y.}~\bibnamefont {Li}},\ }\href {\doibase 10.1088/1751-8113/44/37/375301} {\bibfield  {journal} {\bibinfo  {journal}
  {J. Phys. A: Math. Theor.}\ }\textbf {\bibinfo {volume} {44}},\ \bibinfo
  {pages} {375301} (\bibinfo {year} {2011})}\BibitemShut {NoStop}%
\bibitem [{\citenamefont {Ali}\ \emph {et~al.}(2010{\natexlab{a}})\citenamefont
  {Ali}, \citenamefont {Rau},\ and\ \citenamefont {Alber}}]{ARA10}%
  \BibitemOpen
  \bibfield  {author} {\bibinfo {author} {\bibfnamefont {M.}~\bibnamefont
  {Ali}}, \bibinfo {author} {\bibfnamefont {A.~R.~P.}\ \bibnamefont {Rau}}, \
  and\ \bibinfo {author} {\bibfnamefont {G.}~\bibnamefont {Alber}},\ }\href
  {\doibase 10.1103/PhysRevA.81.042105} {\bibfield  {journal} {\bibinfo
  {journal} {Phys. Rev. A}\ }\textbf {\bibinfo {volume} {81}},\ \bibinfo
  {pages} {042105} (\bibinfo {year} {2010}{\natexlab{a}})}\BibitemShut
  {NoStop}%
\bibitem [{\citenamefont {Ali}\ \emph {et~al.}(2010{\natexlab{b}})\citenamefont
  {Ali}, \citenamefont {Rau},\ and\ \citenamefont {Alber}}]{ARA10E}%
  \BibitemOpen
  \bibfield  {author} {\bibinfo {author} {\bibfnamefont {M.}~\bibnamefont
  {Ali}}, \bibinfo {author} {\bibfnamefont {A.~R.~P.}\ \bibnamefont {Rau}}, \
  and\ \bibinfo {author} {\bibfnamefont {G.}~\bibnamefont {Alber}},\ }\href
  {\doibase 10.1103/PhysRevA.82.069902} {\bibfield  {journal} {\bibinfo
  {journal} {Phys. Rev. A}\ }\textbf {\bibinfo {volume} {82}},\ \bibinfo
  {pages} {069902} (\bibinfo {year} {2010}{\natexlab{b}})}\BibitemShut
  {NoStop}%
\bibitem [{\citenamefont {Lu}\ \emph {et~al.}(2011)\citenamefont {Lu},
  \citenamefont {Ma}, \citenamefont {Xi},\ and\ \citenamefont {Wang}}]{LMXW11}%
  \BibitemOpen
  \bibfield  {author} {\bibinfo {author} {\bibfnamefont {X.-M.}\ \bibnamefont
  {Lu}}, \bibinfo {author} {\bibfnamefont {J.}~\bibnamefont {Ma}}, \bibinfo
  {author} {\bibfnamefont {Z.}~\bibnamefont {Xi}}, \ and\ \bibinfo {author}
  {\bibfnamefont {X.}~\bibnamefont {Wang}},\ }\href {\doibase 10.1103/PhysRevA.83.012327} {\bibfield  {journal} {\bibinfo  {journal} {Phys.
  Rev. A}\ }\textbf {\bibinfo {volume} {83}},\ \bibinfo {pages} {012327}
  (\bibinfo {year} {2011})}\BibitemShut {NoStop}%
\bibitem [{\citenamefont {Chen}\ \emph {et~al.}(2011)\citenamefont {Chen},
  \citenamefont {Zhang}, \citenamefont {Yu}, \citenamefont {Yi},\ and\
  \citenamefont {Oh}}]{CZY+11}%
  \BibitemOpen
  \bibfield  {author} {\bibinfo {author} {\bibfnamefont {Q.}~\bibnamefont
  {Chen}}, \bibinfo {author} {\bibfnamefont {C.}~\bibnamefont {Zhang}},
  \bibinfo {author} {\bibfnamefont {S.}~\bibnamefont {Yu}}, \bibinfo {author}
  {\bibfnamefont {X.~X.}\ \bibnamefont {Yi}}, \ and\ \bibinfo {author}
  {\bibfnamefont {C.~H.}\ \bibnamefont {Oh}},\ }\href {\doibase 10.1103/PhysRevA.84.042313} {\bibfield  {journal} {\bibinfo  {journal} {Phys.
  Rev. A}\ }\textbf {\bibinfo {volume} {84}},\ \bibinfo {pages} {042313}
  (\bibinfo {year} {2011})}\BibitemShut {NoStop}%
\bibitem [{\citenamefont {Huang}(2013{\natexlab{b}})}]{Hua13}%
  \BibitemOpen
  \bibfield  {author} {\bibinfo {author} {\bibfnamefont {Y.}~\bibnamefont
  {Huang}},\ }\href {\doibase 10.1103/PhysRevA.88.014302} {\bibfield  {journal}
  {\bibinfo  {journal} {Phys. Rev. A}\ }\textbf {\bibinfo {volume} {88}},\
  \bibinfo {pages} {014302} (\bibinfo {year} {2013}{\natexlab{b}})}\BibitemShut
  {NoStop}%
\bibitem [{\citenamefont {Luther}\ and\ \citenamefont {Peschel}(1975)}]{LP75}%
  \BibitemOpen
  \bibfield  {author} {\bibinfo {author} {\bibfnamefont {A.}~\bibnamefont
  {Luther}}\ and\ \bibinfo {author} {\bibfnamefont {I.}~\bibnamefont
  {Peschel}},\ }\href {\doibase 10.1103/PhysRevB.12.3908} {\bibfield  {journal}
  {\bibinfo  {journal} {Phys. Rev. B}\ }\textbf {\bibinfo {volume} {12}},\
  \bibinfo {pages} {3908} (\bibinfo {year} {1975})}\BibitemShut {NoStop}%
\bibitem [{\citenamefont {Lukyanov}\ and\ \citenamefont
  {Zamolodchikov}(1997)}]{Luk97}%
  \BibitemOpen
  \bibfield  {author} {\bibinfo {author} {\bibfnamefont {S.}~\bibnamefont
  {Lukyanov}}\ and\ \bibinfo {author} {\bibfnamefont {A.}~\bibnamefont
  {Zamolodchikov}},\ }\href {\doibase 10.1016/S0550-3213(97)00123-5} {\bibfield
   {journal} {\bibinfo  {journal} {Nucl. Phys. B}\ }\textbf {\bibinfo {volume}
  {493}},\ \bibinfo {pages} {571 } (\bibinfo {year} {1997})}\BibitemShut
  {NoStop}%
\bibitem [{\citenamefont {Hikihara}\ and\ \citenamefont
  {Furusaki}(1998)}]{HF98}%
  \BibitemOpen
  \bibfield  {author} {\bibinfo {author} {\bibfnamefont {T.}~\bibnamefont
  {Hikihara}}\ and\ \bibinfo {author} {\bibfnamefont {A.}~\bibnamefont
  {Furusaki}},\ }\href {\doibase 10.1103/PhysRevB.58.R583} {\bibfield
  {journal} {\bibinfo  {journal} {Phys. Rev. B}\ }\textbf {\bibinfo {volume}
  {58}},\ \bibinfo {pages} {R583} (\bibinfo {year} {1998})}\BibitemShut
  {NoStop}%
\bibitem [{\citenamefont {Lukyanov}(1999)}]{Luk99}%
  \BibitemOpen
  \bibfield  {author} {\bibinfo {author} {\bibfnamefont {S.}~\bibnamefont
  {Lukyanov}},\ }\href {\doibase 10.1103/PhysRevB.59.11163} {\bibfield
  {journal} {\bibinfo  {journal} {Phys. Rev. B}\ }\textbf {\bibinfo {volume}
  {59}},\ \bibinfo {pages} {11163} (\bibinfo {year} {1999})}\BibitemShut
  {NoStop}%
\bibitem [{\citenamefont {Affleck}(1998)}]{Aff98}%
  \BibitemOpen
  \bibfield  {author} {\bibinfo {author} {\bibfnamefont {I.}~\bibnamefont
  {Affleck}},\ }\href {\doibase 10.1088/0305-4470/31/20/002} {\bibfield
  {journal} {\bibinfo  {journal} {J. Phys. A: Math. Gen.}\ }\textbf {\bibinfo
  {volume} {31}},\ \bibinfo {pages} {4573} (\bibinfo {year}
  {1998})}\BibitemShut {NoStop}%
\bibitem [{\citenamefont {Lukyanov}(1998)}]{Luk98}%
  \BibitemOpen
  \bibfield  {author} {\bibinfo {author} {\bibfnamefont {S.}~\bibnamefont
  {Lukyanov}},\ }\href {\doibase 10.1016/S0550-3213(98)00249-1} {\bibfield
  {journal} {\bibinfo  {journal} {Nucl. Phys. B}\ }\textbf {\bibinfo {volume}
  {522}},\ \bibinfo {pages} {533 } (\bibinfo {year} {1998})}\BibitemShut
  {NoStop}%
\bibitem [{\citenamefont {Lieb}\ \emph {et~al.}(1961)\citenamefont {Lieb},
  \citenamefont {Schultz},\ and\ \citenamefont {Mattis}}]{LSM61}%
  \BibitemOpen
  \bibfield  {author} {\bibinfo {author} {\bibfnamefont {E.}~\bibnamefont
  {Lieb}}, \bibinfo {author} {\bibfnamefont {T.}~\bibnamefont {Schultz}}, \
  and\ \bibinfo {author} {\bibfnamefont {D.}~\bibnamefont {Mattis}},\ }\href
  {\doibase 10.1016/0003-4916(61)90115-4} {\bibfield  {journal} {\bibinfo
  {journal} {Ann. Phys. (N.Y.)}\ }\textbf {\bibinfo {volume} {16}},\ \bibinfo
  {pages} {407 } (\bibinfo {year} {1961})}\BibitemShut {NoStop}%
\bibitem [{\citenamefont {McCoy}(1968)}]{Mcc68}%
  \BibitemOpen
  \bibfield  {author} {\bibinfo {author} {\bibfnamefont {B.~M.}\ \bibnamefont
  {McCoy}},\ }\href {\doibase 10.1103/PhysRev.173.531} {\bibfield  {journal}
  {\bibinfo  {journal} {Phys. Rev.}\ }\textbf {\bibinfo {volume} {173}},\
  \bibinfo {pages} {531} (\bibinfo {year} {1968})}\BibitemShut {NoStop}%
\bibitem [{\citenamefont {Pfeuty}(1970)}]{Pfe70}%
  \BibitemOpen
  \bibfield  {author} {\bibinfo {author} {\bibfnamefont {P.}~\bibnamefont
  {Pfeuty}},\ }\href {\doibase 10.1016/0003-4916(70)90270-8} {\bibfield
  {journal} {\bibinfo  {journal} {Ann. Phys. (N.Y.)}\ }\textbf {\bibinfo
  {volume} {57}},\ \bibinfo {pages} {79 } (\bibinfo {year} {1970})}\BibitemShut
  {NoStop}%
\bibitem [{\citenamefont {Niemeijer}(1967)}]{Nie67}%
  \BibitemOpen
  \bibfield  {author} {\bibinfo {author} {\bibfnamefont {T.}~\bibnamefont
  {Niemeijer}},\ }\href {\doibase 10.1016/0031-8914(67)90235-2} {\bibfield
  {journal} {\bibinfo  {journal} {Physica}\ }\textbf {\bibinfo {volume} {36}},\
  \bibinfo {pages} {377 } (\bibinfo {year} {1967})}\BibitemShut {NoStop}%
\bibitem [{\citenamefont {Barouch}\ and\ \citenamefont {McCoy}(1971)}]{BM71}%
  \BibitemOpen
  \bibfield  {author} {\bibinfo {author} {\bibfnamefont {E.}~\bibnamefont
  {Barouch}}\ and\ \bibinfo {author} {\bibfnamefont {B.~M.}\ \bibnamefont
  {McCoy}},\ }\href {\doibase 10.1103/PhysRevA.3.786} {\bibfield  {journal}
  {\bibinfo  {journal} {Phys. Rev. A}\ }\textbf {\bibinfo {volume} {3}},\
  \bibinfo {pages} {786} (\bibinfo {year} {1971})}\BibitemShut {NoStop}%
\end{thebibliography}
\end{document}